\documentclass[apex,twocolumn]{jjap3}
\usepackage{graphicx}
\usepackage{array}
\usepackage{txfonts}

\title{Cubic C$_{20}$: An intrinsic superconducting carbon allotrope}
\author{Ying Yu$^{1}$, Xun-Wang Yan$^{2}$, Fengjie Ma$^{3}$, Miao Gao$^{1,4}$\thanks{gaomiao@nbu.edu.cn}, and Zhong-Yi Lu$^{5}$}

\inst{$^{1}$Department of Physics, School of Physical Science and Technology, Ningbo University, Zhejiang 315211, China \\
$^{2}$College of Physics and Engineering, Qufu Normal University, Shandong 273165, China \\
$^{3}$The Center for Advanced Quantum Studies and Department of Physics, Beijing Normal University, Beijing 100875, China \\
$^{4}$School of Physics, Zhejiang University, Hangzhou 310058, China \\
$^{5}$Department of Physics, Renmin University of China, Beijing 100872, China}

\abst{Finding intrinsic carbon superconductor is an interesting topic. Based on density functional first-principles calculations, we first study the phonon-mediated superconductivity in
a cubic metallic carbon allotrope, namely sc-C$_{20}$, which has been synthesized in experiment.
The electron-phonon coupling is accurately computed with Wannier interpolation method. By solving the Eliashberg equations, we predict that
sc-C$_{20}$ is an intrinsic carbon superconductor, without introducing any guest atoms or doping, whose transition temperature is determined to be about 24 K.
Our findings enrich the family of carbon-based superconductors.}

\begin{document}

\maketitle

The studies of carbon-based superconductor can be traced back to the 1960s.
Superconductivity was first discovered in graphite intercalation compounds (GICs) below 1 K, such as KC$_8$ \cite{Hannay-PRL14}.
Under ambient pressure, the highest $T_c$ for GICs is 11.5 K in CaC$_6$ \cite{Emery-PRL95}.
Graphene, the two-dimensional form of graphite, has gapless Dirac bands.
Superconductivity in doped graphene has been extensively investigated \cite{Uchoa-PRL98,Profeta-NP8,Chapman-SR6}.
Recently, correlated insulating states were observed in twisted bilayer graphene, and superconductivity emerges at 1.7 K after electrostatic doping \cite{Cao-Nature556-1,Cao-Nature556-2}.
For twisted trilayer graphene, rich phase diagram and better tunability of electric field were realized, compared with the bilayer case \cite{Zhu-PRL125,Park-N590}, providing a fascinating playground to explore the interplay between correlated states and superconductivity.
Insulator to superconductor transition was reported in boron-doped diamond \cite{Ekimov-Nature428},
in which the $T_c$ exhibits a positive dependence on the proportion of boron that incorporated into diamond \cite{Takano-DRM16}.
It was found that the onset $T_c$ of 27\% boron-doped Q-carbon is 55 K \cite{Bhaumik-AN11}.
Fullerene shows superconductivity at 18 K, after the potassium intercalation with stoichiometry of K$_3$C$_{60}$ \cite{Hebard-Nature350}.
Subsequently, the $T_c$ was improved to 33 K in RbCs$_2$C$_{60}$ \cite{Tanigaki-Nature352} and 40 K in Cs$_3$C$_{60}$ \cite{Palstra-SSC93}. Signatures of superconductivity were also detected in single- and multi-walled carbon nanotubes \cite{Kociak-PRL86,Takesue-PRL96}.
However, the superconductivity transition is quite sensitive to the configuration of Au electrode \cite{Takesue-PRL96}.
Organic compounds are possible candidates of carbon-based superconductors, such as K$_{3.3}$Picene \cite{Mitsuhashi-Nature464}, but controversy still exists \cite{Heguri-PRB92}.

\begin{figure}[tbh]
\centering
\includegraphics[width=8.6cm]{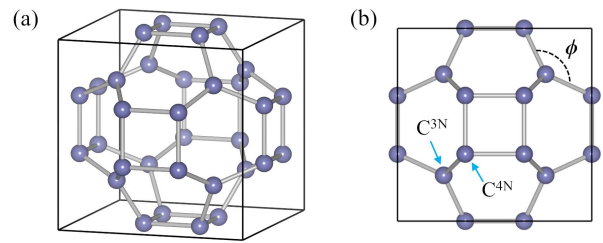}
\caption{(a) Three-dimensional view of the ball-stick model of sc-C$_{20}$. (b) View along the [001] direction. The solid black lines represent the cubic unit cell. C$^\text{3N}$ and C$^\text{4N}$
denote two nonequivalent carbon atoms. $\phi$ is the bond angle among $sp^2$-bonded atoms. The ball-stick models are plotted with VESTA \cite{Momma-JAC44}.}
\label{fig:Stru}
\end{figure}

As what mentioned above, introducing guest atom or doping is inescapable to realize superconductivity in carbon materials.
Therefore, it is quite interesting to know whether there is a carbon allotrope showing intrinsic superconductivity.
To achieve this purpose, the first step is finding metallic carbon allotropes. K. Yamada prepared a cubic modification of carbon by shock compression method, with carbon black and tetracyanoethylene as starting materials \cite{Yamada-Carbon41}. In the theoretical side, a simple cubic carbon, namely sc-C$_{20}$ [Fig.~\ref{fig:Stru}], was proposed \cite{Ribeiro-PRB74}.
The lattice constant of sc-C$_{20}$ is in excellent agreement with the synthesized cubic carbon. Moreover, by comparing the X-ray diffraction (XRD) patterns, it was found that most XRD peaks from the experiment can be interpreted through sc-C$_{20}$ \cite{He-Carbon112}. Band structure calculation indicated that sc-C$_{20}$ is metallic \cite{Ribeiro-PRB74}.
Besides sc-C$_{20}$, several metallic carbon allotropes were proposed theoretically, such as bct-4 \cite{Hoffman-JACS105}, T6-carbon \cite{Zhang-PNAS110}, Hex-C$_{24}$ \cite{Bu-JMCC2}, H$_{18}$ carbon \cite{Zhao-SC6}, and C14-diamond \cite{Wu-Carbon123}.
However, the electron-phonon coupling (EPC) and potential phonon-mediated superconductivity in all these metallic carbon allotropes have never been investigated so far.
Two hypothetical $sp^2$-hybridized carbons, namely GT-8 and GT-16 were suggested to be superconducting, with $T_c$ being 5.2 K and 14.0 K, respectively \cite{Hu-PCCP17}.
However, the evidences about their syntheses are not available.
Very recently, different structural models were adopted to study the EPC and superconductivity of graphite-diamond hybrid, with predicted $T_c$ ranging from 2 K to 42 K \cite{Ge-MTP23}.
Thus, the phase space of superconducting carbon allotrope needs to be further enriched.

Considering the fact that sc-C$_{20}$ has been synthesized, we perform the first-principles calculations of EPC and phonon-mediated superconductivity in sc-C$_{20}$.
The atomic structure, elastic properties, electronic structure, phonons, isotropic Eliashberg spectral function, and evolution of superconducting gap versus temperature are systematically investigated in this work.
As revealed by our calculations, the metallicity of sc-C$_{20}$ originates from the $p$ orbitals of carbon atoms, which are perpendicular to the $sp^2$-bonding planes. Phonon modes, including $A_{2u}$, $T_{2g}$, $A_{1g}$, $E_g$, and $A_{2g}$, involve in the EPC, providing multiple channels to pair electrons.
The EPC constant $\lambda$ and $\omega_\text{log}$ are equal to 0.63 and 60.81 meV, respectively, giving rise to superconductivity with $T_c$ being about 24 K.
As a comparison, the EPC of other five carbon allotropes are also studied, including
bct-4, T6-carbon, Hex-C$_{24}$, H$_{18}$ carbon, and C14-diamond. None of these compounds show superconductivity above 8 K.

\begin{figure}[b]
\centering
\includegraphics[width=8.6cm]{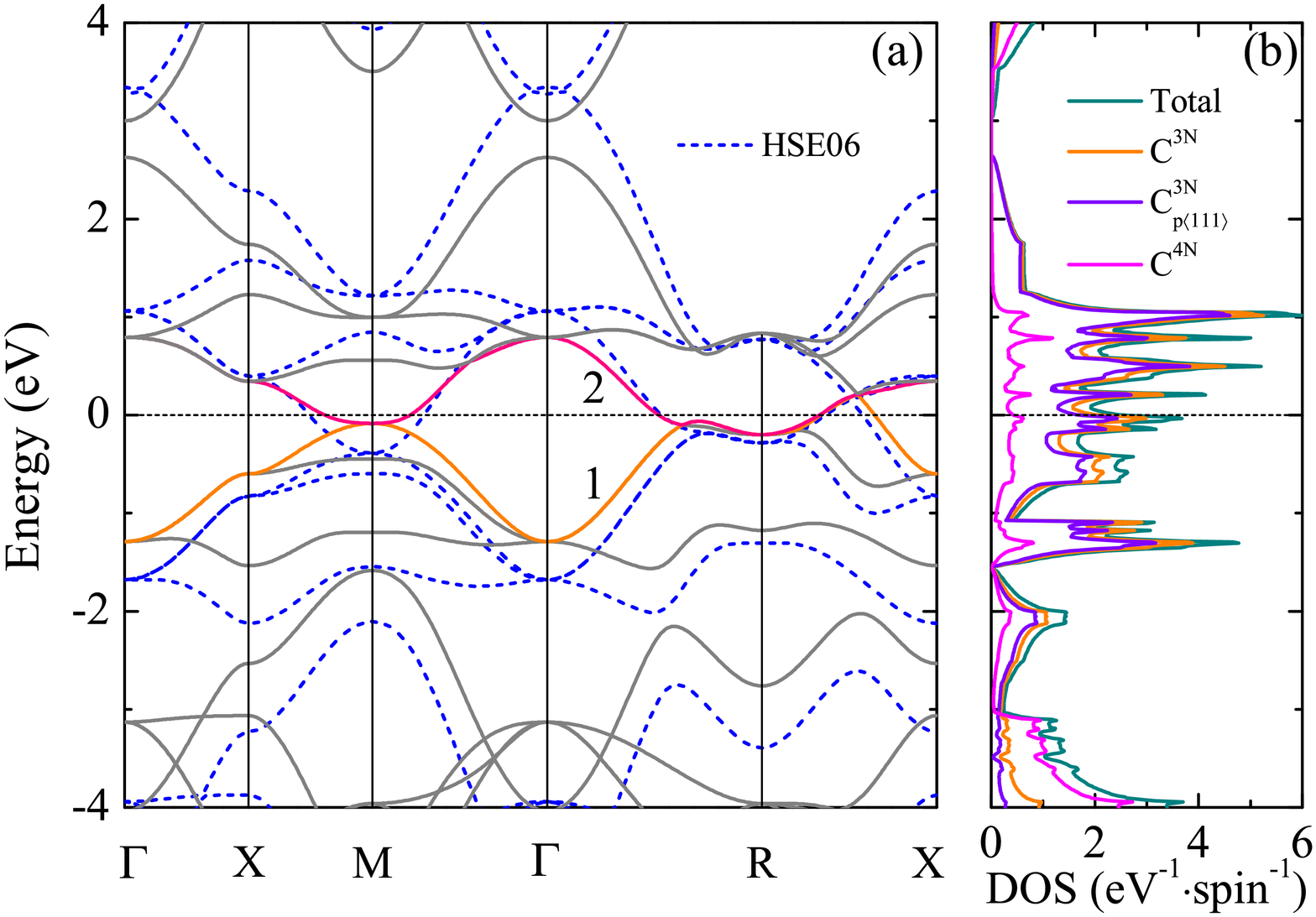}
\includegraphics[width=8.6cm]{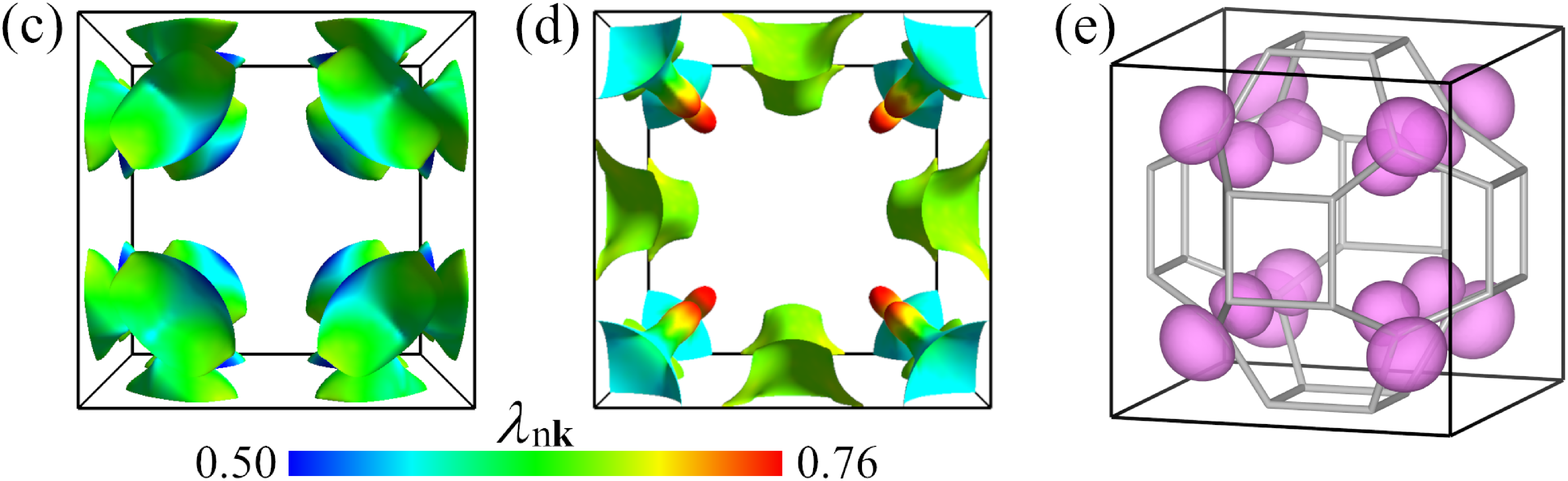}
\caption{Electronic structure for sc-C$_{20}$. (a) Band structure. The solid lines denote the PBE bands. The partially filled bands labeled as 1 and 2, are presented with different colors. The Fermi level is set to zero. (b) Total and partial density of states generated by PBE functional. (c)-(d) Fermi surfaces corresponding to band 1 and band 2, whereas the
distribution of {\bf k}-resolved EPC constant $\lambda_{n{\bf k}}$ is shown. The FermiSurfer package is adopted \cite{Kawamura-CPC239}.
(e) Isosurface of charge density of strongly coupled electronic Kohn-Sham states drawn with VESTA \cite{Momma-JAC44}. State at {\bf k} point ($\pm$0.26, $\pm$0.26, $\pm$0.26) are included. The isovalue is set to 0.025 $e$/Bohr$^3$.}
\label{fig:Band}
\end{figure}

Our density functional theory calculations were carried out based on the plane wave basis and pseudopotential methods.
The Quantum-ESPRESSO package was adopted \cite{pwscf}.
We calculated the electronic states and phonon perturbation potentials \cite{Giustino-PRB76}
using the generalized gradient approximation of Perdew-Burke-Ernzerhoff (PBE)
formula \cite{Perdew-PRL77} and the optimized norm-conserving Vanderbilt pseudopotentials \cite{ONCV-PP}.
The kinetic energy
cut-off and the charge density cut-off were set to 80 Ry and 320 Ry, respectively.
A {\bf k} mesh of 18$\times $18$\times $18 points in combination with
a Methfessel-Paxton smearing \cite{Methfessel-PRB40} of 0.02 Ry, was employed to calculate the self-consistent charge densities.
Within the framework of density-functional perturbation theory \cite{Baroni-RMP73}, we computed the dynamical matrices and the perturbation potentials
on a 6$\times $6$\times $6 mesh.
To construct the maximally localized Wannier functions of sc-C$_{20}$ \cite{Pizzi-JPCM32}, we chose 36 hybridized $\sigma$ states localized in the middle of carbon-carbon bonds and
8 $s$-type functions at the $sp^2$-bonded carbon sites.
The convergence of EPC constant $\lambda$ was extensively checked through
fine electron (60$\times $60$\times $60) and phonon (20$\times $20$\times $20) grids with the Electron-Phonon Wannier code \cite{Ponce-CPC209}.
For the energy bands consistency between Wannier interpolation and first-principles calculations,
the convergence test of EPC constant $\lambda$, and the spatial decay of electronic Hamiltonian in Wannier representation, please see the supplemental material \cite{supp}.
The Dirac
$\delta $-functions for electrons and phonons were smeared out by a Gaussian function with the
widths of 20 meV and 0.5 meV, respectively. We determined the $T_c$ by solving the isotropic Eliashberg equations.
The sum over Matsubara frequencies was truncated with $\omega_c$=16 eV, about 100 times that of the highest phonon frequency.

The crystal structure of sc-C$_{20}$ is shown in Fig. \ref{fig:Stru}. Carbon atoms can be classified into two types. Eight $sp^2$-bonded carbon atoms are seated close to
the cell corners. Twelve $sp^3$-hybridized carbon atoms locate around the face centers of the cubic cell, forming squares. According to the number of the nearest neighbors, these two nonequivalent carbon atoms are denoted as C$^\text{3N}$ and C$^\text{4N}$, respectively [Fig. \ref{fig:Stru}(b)].
sc-C$_{20}$ belongs to the space group $Pm\bar{3}m$ (No. 221).
After optimization, the lattice constant of sc-C$_{20}$ is found to be 5.2135 {\AA}, slightly larger than the one obtained by local density approximation (LDA) \cite{Ribeiro-PRB74,He-Carbon112}.
C$^\text{3N}$ and C$^\text{4N}$ occupy 8$g$ (0.2376, 0.2376, 0.2376) and 12$i$ (0.0000, 0.3485, 0.3485) Wyckoff positions.
The bond angle $\phi$ is 119.9$^\circ$ [Fig. \ref{fig:Stru}(b)], indicating a perfect $sp^2$ hybridization, like in graphene.
The mechanical stability of sc-C$_{20}$ is examined by calculating the elastic constants $C_{ij}$.
For cubic crystals, there are only three independent $C_{ij}$ coefficients, namely $C_{11}$, $C_{12}$, and $C_{44}$,
which are determined to be 537.0 GPa, 224.2 GPa, and 228.1 GPa, respectively. It is known that the mechanical stability conditions for cubic systems are
$C_{11}-C_{12}>0$, $C_{11}+2C_{12}>0$, and $C_{44}>0$ \cite{Boron-1954}, which are satisfied in sc-C$_{20}$. Using Voigt-Reuss-Hill approximation \cite{Hill-PPSA65}, the bulk modulus $B$, shear modulus $G$, Young modulus $E$, and Poisson's ratio $\nu$ are 328.5 GPa, 196.1 GPa, 490.7 GPa, and 0.25, respectively.
The elastic constants and bulk modulus are relatively smaller with respect to the LDA results \cite{Ribeiro-PRB74,He-Carbon112}, due to the well-known overbinding problem of LDA.
We also estimate the Vickers hardness of sc-C$_{20}$ with Chen's model \cite{Chen-Inter19}, $H_v$=$2(k^2G)^{0.585}-3$, in which the Pugh modulus ratio $k$ equals $G/B$. sc-C$_{20}$ is not a superhard material with $H_v$ of 21.0 GPa.

\begin{figure}[b]
\centering
\includegraphics[width=8.6cm]{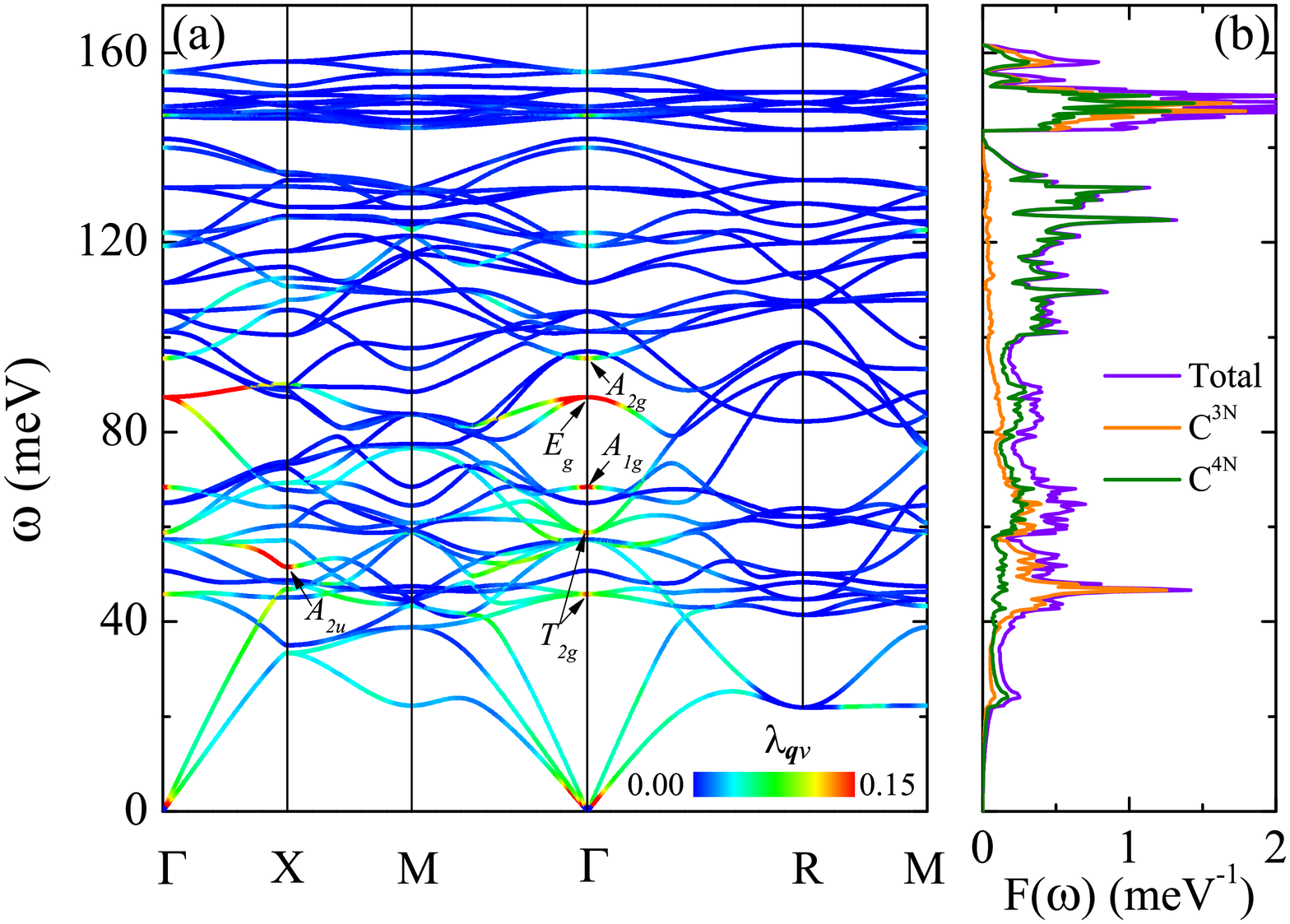}
\caption{Lattice dynamics of sc-C$_{20}$. (a) Phonon spectrum, colored by EPC constant $\lambda_{{\bf q}v}$. (b) Total and projected phonon DOS $F(\omega)$.}
\label{fig:phonon}
\end{figure}

The electronic structure of sc-C$_{20}$ is presented in Fig. \ref{fig:Band}. It is clear that sc-C$_{20}$ is a metal, with two energy bands across the Fermi level [Fig. \ref{fig:Band}(a)].
Since PBE functional trends to underestimate the band gap, we also calculate the band structure using the hybrid functional HSE06 \cite{Heyd-JCP124}.
Although, a repulsion effect is observed in comparison with the PBE bands. The HSE06 results resemble those of the PBE around the Fermi level.
According to the partial density of states (DOS), C$^\text{3N}$ atoms dominate the DOS from -3.0 eV to 3.0 eV [Fig. \ref{fig:Band}(b)]. Specifically, they contribute about 81.5\% of the DOS at the Fermi level, i.e. $N(0)$.
The $p$ orbitals of C$^\text{3N}$ that perpendicular to the $sp^2$-bonding plane are along the $\langle111\rangle$ direction, which are denoted as C$_{p\langle111\rangle}^{\text{3N}}$.
In particular, the DOS of C$^\text{3N}$ almost originates from the C$_{p\langle111\rangle}^{\text{3N}}$ orbitals.
Nevertheless, contribution of C$^\text{4N}$ atoms mainly lies below -3.0 eV. The Fermi surfaces are shown in Fig. \ref{fig:Band}(c) and Fig. \ref{fig:Band}(d), with color mapping of the EPC strength $\lambda_{n{\bf k}}$. All the pieces of the Fermi surfaces are close to the boundaries of Brillouin zone. The hotspots in Fig. \ref{fig:Band}(d) stand for the
Kohn-Sham states that possess the strongest coupling with phonons. As confirmed by the charge distribution, these strongly coupled electronic states just correspond to the C$_{p\langle111\rangle}^{\text{3N}}$ orbitals [Fig. \ref{fig:Band}(e)].

Figure \ref{fig:phonon} shows the phonon spectrum and phonon DOS.
Beside mechanical stability, sc-C$_{20}$ is proved to be dynamically stable, as indicated by the phonon spectrum, whereas no imaginary phonon modes appear [Fig. \ref{fig:phonon}(a)].
The symmetries of phonon modes which strongly couple with electrons are marked. For instance, the $A_{2u}$ mode at $X$ point, $T_{2g}$, $A_{1g}$, $E_g$, and $A_{2g}$ modes at $\Gamma$ point.
The highest peak of phonon DOS is around 150 meV, originating from dispersionless modes [Fig. \ref{fig:phonon}(b)]. These modes mix the movements of C$^\text{4N}$ and C$^\text{3N}$ atoms together, as demonstrated by the projected phonon DOS. The vibration of C$^\text{4N}$ holds a dominant position from 80 meV to 140 meV.
The situation is reversed below 60 meV.

\begin{figure}[tbh]
\centering
\includegraphics[width=8.6cm]{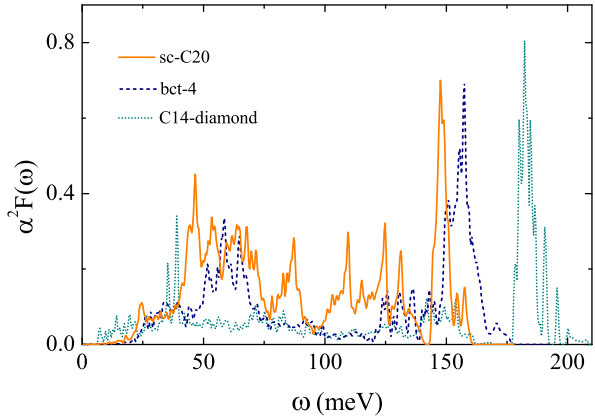}
\caption{Eliashberg spectral functions $\alpha^2F(\omega)$ for sc-C$_{20}$, bct-4, and C14-diamond. The results of T6, Hex-C$_{24}$, and H$_{18}$ carbon are not shown, due to weak EPC.}
\label{fig:a2f}
\end{figure}

The Eliashberg spectral functions $\alpha^2F(\omega)$ of sc-C$_{20}$, bct-4, and C14-diamond are given in Fig. \ref{fig:a2f}. The multi-peak behavior of $\alpha^2F(\omega)$ in sc-C$_{20}$ naturally reflects the existence of various
strongly coupled phonon modes, as uncovered in Fig. \ref{fig:phonon} (a). The peak near 150 meV is resulted from the large phonon DOS. In contrast, the reduced peak around 87 meV can be attributed to the limited DOS, although the $E_g$ modes at $\Gamma$ point have sizeable EPC strength. Compared with bct-4 and C14-diamond, two important features of
$\alpha^2F(\omega)$ are revealed for sc-C$_{20}$. On one hand, the phonons of sc-C$_{20}$ exhibit obvious softening, with the highest phonon frequency being about 160 meV, which is reduced by 23.0\% with respect to that in C14-dimaond. On the other hand, the spectral weight of $\alpha^2F(\omega)$ of sc-C$_{20}$ in the intermediate frequency range is significant, for example, from 80 meV to 120 meV. These two factors play vital roles in the enhancement of EPC in sc-C$_{20}$.
The EPC constant $\lambda$ and logarithmic average frequency $\omega_\text{log}$ can be acquired through
$\lambda=2\int\frac{\alpha^2F(\omega)}{\omega}d\omega$, and $\omega_{\text{log}}=\exp\Big[\frac{2}{\lambda}\int\frac{d\omega}{\omega}\alpha^2F(\omega)\ln\omega\Big]$.
We find that the EPC constants $\lambda$ are 0.63, 0.42, 0.37, 0.16, 0.07, and 0.21 in sc-C$_{20}$, bct-4, C14-diamond, T6-carbon, Hex-C$_{24}$, and H$_{18}$ carbon, respectively.
The logarithmic average frequencies $\omega_\text{log}$ are computed to be 60.81, 66.28, 47.10, 88.23, 111.60, and 72.18 meV, correspondingly.

\begin{figure}[t]
\begin{center}
\includegraphics[width=8.6cm]{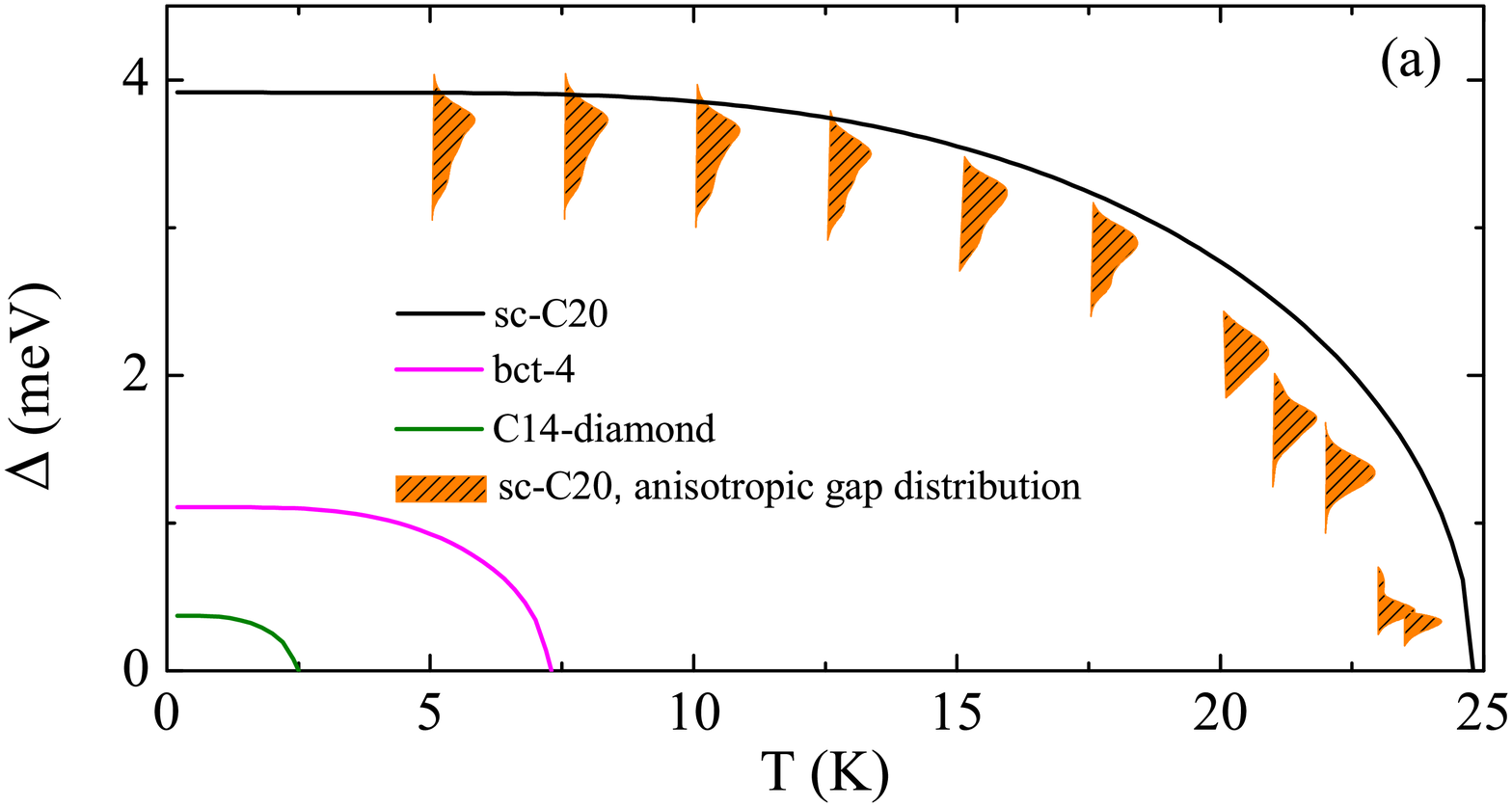}
\includegraphics[width=8.6cm]{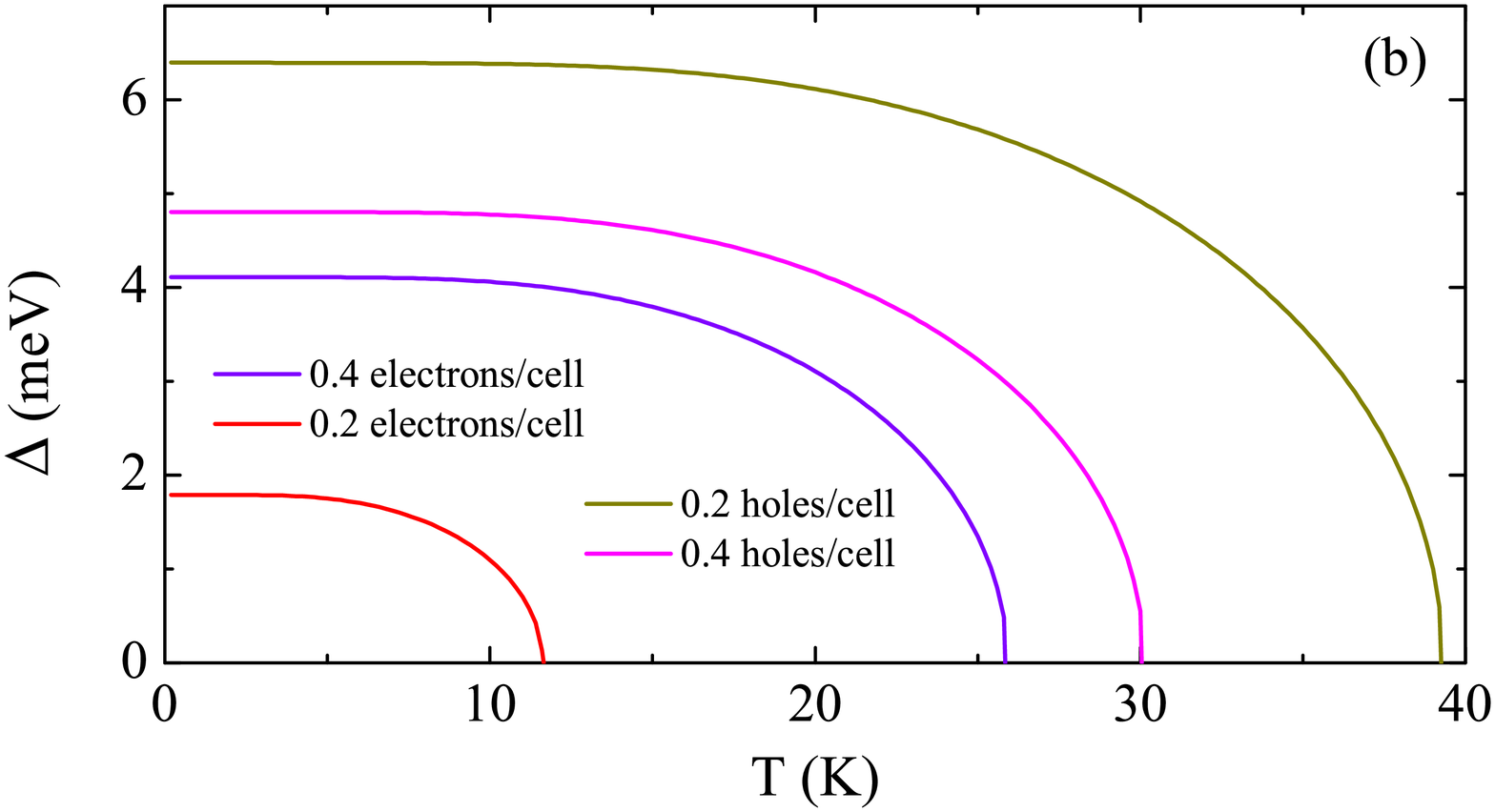}
\caption{(a) Superconducting energy gaps versus temperature for sc-C$_{20}$, bct-4, and C14-diamond, according to the results of isotropic Eliashberg equations.
The shadowed yellow curve is the distribution of normalized superconducting energy gap generated by using the anisotropic Eliashberg equations.
(b) $T_c$ of doped sc-C$_{20}$.
In our calculations, the screened Coulomb potential $\mu^*$ is set to 0.10.}
\label{fig:Tc}
\end{center}
\end{figure}

In order to determine the $T_c$, the isotropic Eliashberg equations are solved self-consistently \cite{Eliashberg-ZETF966,Margine-PRB87}.
The transition temperatures are calculated to be 24.6 K, 7.2 K, and 2.4 K, for sc-C$_{20}$, bct-4, and C14-diamond [Fig.~\ref{fig:Tc}(a)].
Close to zero temperature, the superconducting energy gaps are equal to 3.91, 1.11, and 0.37 meV, respectively.
No superconducting transition occurs in T6-carbon, Hex-C$_{24}$, and H$_{18}$ carbon.
We also take the anisotropy of EPC into consideration. The superconducting energy gaps show visible anisotropy in sc-C$_{20}$.
But the gap values group together, reflecting a single-gap nature. This can be rationalized by the fact that only one type orbital dominates the $N(0)$, i.e., C$_{p\langle111\rangle}^{\text{3N}}$.
Moreover, the $T_c$ is computed as 23.5 K, almost unaffected compared with the isotropic case. Considering the multiple-peak structure of DOS near the Fermi level [Fig.~\ref{fig:Band}(b)], and the sensitivity of $T_c$ to $N(0)$,
we further calculate the $T_c$ of doped sc-C$_{20}$ to investigate its robustness against the change of $N(0)$. The doping concentrations are set to 0.4 electrons/cell, 0.2 electrons/cell, 0.2 hole/cell, and 0.4 hole/cell, respectively. Accordingly, we find that $N(0)$ equal to 3.34, 1.97, 3.46, and 3.01 states/spin/eV/cell, for above mentioned four doping cases, in comparison with 2.65 states/spin/eV/cell of pristine sc-C$_{20}$. After solving the isotropic Eliashberg equations, the transition temperatures are determined to be 25.8, 11.6, 39.2, and 30.0 K, respectively [Fig.~\ref{fig:Tc}(b)].
Although 0.2 electrons/cell doping leads to depressed $T_c$, doping holes can markedly boost the $T_c$. Our findings strongly suggest that sc-C$_{20}$ is a unique intrinsic superconducting carbon allotrope, with $T_c$ even higher than most carbon-based superconductors.

In summary, we have investigated the phonon-mediated superconductivity in six carbon allotropes for the first time.
The unpaired $p$ orbitals that pointing to the corners of the cubic cell can account for the metallicity in sc-C$_{20}$.
The symmetries of strongly coupled phonon modes are identified.
Our calculations suggest that sc-C$_{20}$ is an outstanding intrinsic carbon superconductor.
There is no need to introduce guest atoms or doping to realize superconductivity, unlike other carbon-based compounds.
The transition temperature of sc-C$_{20}$ is higher than those in graphite intercalation compounds, and also comparable with respect to that predicted in graphite-diamond hybrid.
The confirmation of its superconductivity in experiment is called for.

\acknowledgment
This work was supported by the National Natural Science Foundation of China (Grant Nos. 11974194, 11974207, 12074040, 11934020, and 11888101).

\end{document}